\documentclass[a4paper,]{jpconf}
\usepackage{graphicx}
\begin{document}
\title{Core-collapse supernovae and neutrino properties}

%dune, scypy ,lesa, SK figure

\author{Maria Cristina Volpe}

\address{Université Paris Cité, CNRS, Astroparticule et Cosmologie, F-75013 Paris, France}

\ead{volpe@apc.in2p3.fr}

\begin{abstract}
We highlight developments in the domain of supernova neutrinos. We discuss the importance of the future observation, by running and upcoming experiments, of the neutrino signals from the next supernova as well as of the diffuse supernova neutrino background.
\end{abstract}

\section{Introduction}
In Nature and with Earth-based experiments we have neutrino sources of all flavors with fluxes that cover about 30 orders of magnitude and range from meV to PeV \cite{IceCube:2014stg} energies. Two neutrino backgrounds have not been observed yet. The cold cosmological one, which decoupled 1s after the Big Bang, left an imprint on primordial abundances of light elements and on large scale structures. Its observation requires new detection methods, such as the capture on radioactive nuclei, a process without threshold first proposed by Weinberg \cite{Weinberg:1962zza}. Revived by Cocco, Mangano and Messina \cite{Cocco:2007za} this idea was further studied for example in \cite{Lazauskas:2007da,Long:2014zva,Roulet:2018fyh} and exploited by the PTOLEMY project, currently under study \cite{PTOLEMY:2018jst}. 
     
Core-collapse supernovae are stars with more than 6 solar masses (M$_{\odot}$). During the late stages of their evolution, they develop an O-Ne-Mg 
(6  M$_{\odot} < $  M $ < $ 8 M$_{\odot}$) ) or iron (M $>$ 8 M$_{\odot}$) cores. There are supernovae of type II or I b/c depending on the hydrogen envelope, if it is present or absent. Core-collapse supernovae
undergo gravitational collapse at the end of their life and emit about 10$^{58}$ neutrinos that take away 99 $\%$ of the gravitational binding energy (about $ 3 ~10^{53}$ erg), as first suggested by Colgate and White \cite{Colgate:1966ax}. Only about 1$\%$ corresponds to the explosion kinetic energy whereas circa $0.01 \%$ is taken by photons. Thus supernovae represent one of the most powerful sources of neutrinos of all flavors, during 10 seconds, and constitute a rich laboratory for particle physics and astrophysics.  

So far, the only supernova seen through its neutrinos is SN1987A (Figure 1), located at 50 kpc (163000 light-years) from the Earth, in the Large Magellanic Cloud, a satellite galaxy of the Milky Way. Kamiokande \cite{Hirata:1987hu}, IMB \cite{Bionta:1987qt} and Baksan \cite{Alekseev:1988gp} detectors recorded 24 $\bar{\nu}_e$ events with average energies, time spread and total gravitational energy in agreement with expectations (under the equipartition hypothesis) \cite{Loredo:2001rx,Vissani:2014doa}. A few still debated events were detected 5 hours before the others in the Mont Blanc observatory \cite{Aglietta:1987it}. The time signal supported the delayed neutrino-heating explosion mechanism suggested by Bethe and Wilson \cite{Bethe:1985sox} and provided numerous limits on unknown neutrino properties, non-standard particles like axions and interactions (see for example \cite{Raffelt:1996wa}). 
\begin{figure}
\begin{center}
\includegraphics[scale=0.13]{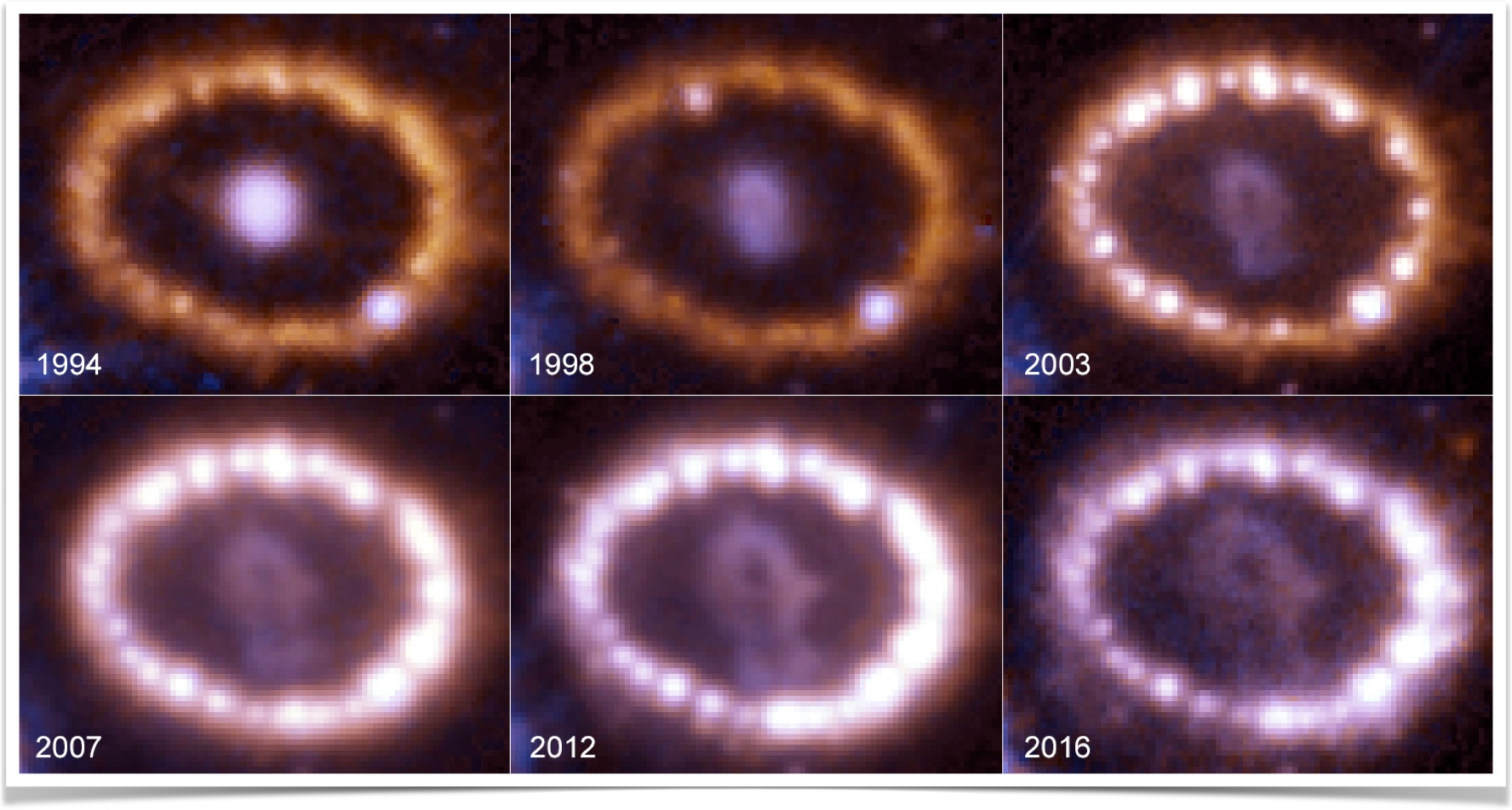}
\end{center}
\caption{\label{label} Hubble Space Telescope images (1994 to 2016) of SN1987A at 50 kpc, in the Large Magellanic Cloud (ESA/Hubble) \cite{HST}.}
\end{figure}

Unfortunately supernovae are rare. In our Galaxy the mean-time for their occurrence  is of 50 $\pm$ 20 years (see \cite{Costantini:2005un} and references therein). At 0.2 kpc a supernova candidate, Betelgeuse. The neutrino luminosity curve of the next supernova, if close enough, will be precisely measured, in all flavors. This awaited observation will bring key information on the longstanding open question of the supernova explosion mechanism (Hoyle and Fowler suggested back in the sixties that the stellar death of massive stars is due to core implosion) as well as on neutrino properties like the neutrino magnetic moment. 

Indeed the strong asphericities and mixing observed in SN1987A ejecta gave momentum to the developments of multidimensional supernova simulations. While two-dimensional simulations show successful (albeit overemphasized) explosions, three-dimensional ones are close to explosion (see for example \cite{Janka:2017vcp,Radice:2017kmj,Bruenn:2018wpz}). The death of massive stars is likely due to an interplay of convection, turbulence, neutrino heating behind the shock and hydrodynamical instabilities (SASI, Figure 2). Interestingly, characteristic imprints are left by the SASI on the time signal, as discussed by several authors (Figure 3). Another hydrodynamical instability termed LESA creates an asymmetric neutrino emission due to convection \cite{Tamborra:2014aua}
($\nu_e$ and $\bar{\nu}_e$ having a dipole pattern, $\nu_x$ being more spherically symmetric).
Thus, as with SN1987A, the current paradigm for supernova explosions could be confirmed/refuted by the observation of the next supernova.

\begin{center}
\begin{figure}[h]
\begin{center}
\begin{minipage}{14pc}
\includegraphics[width=12pc]{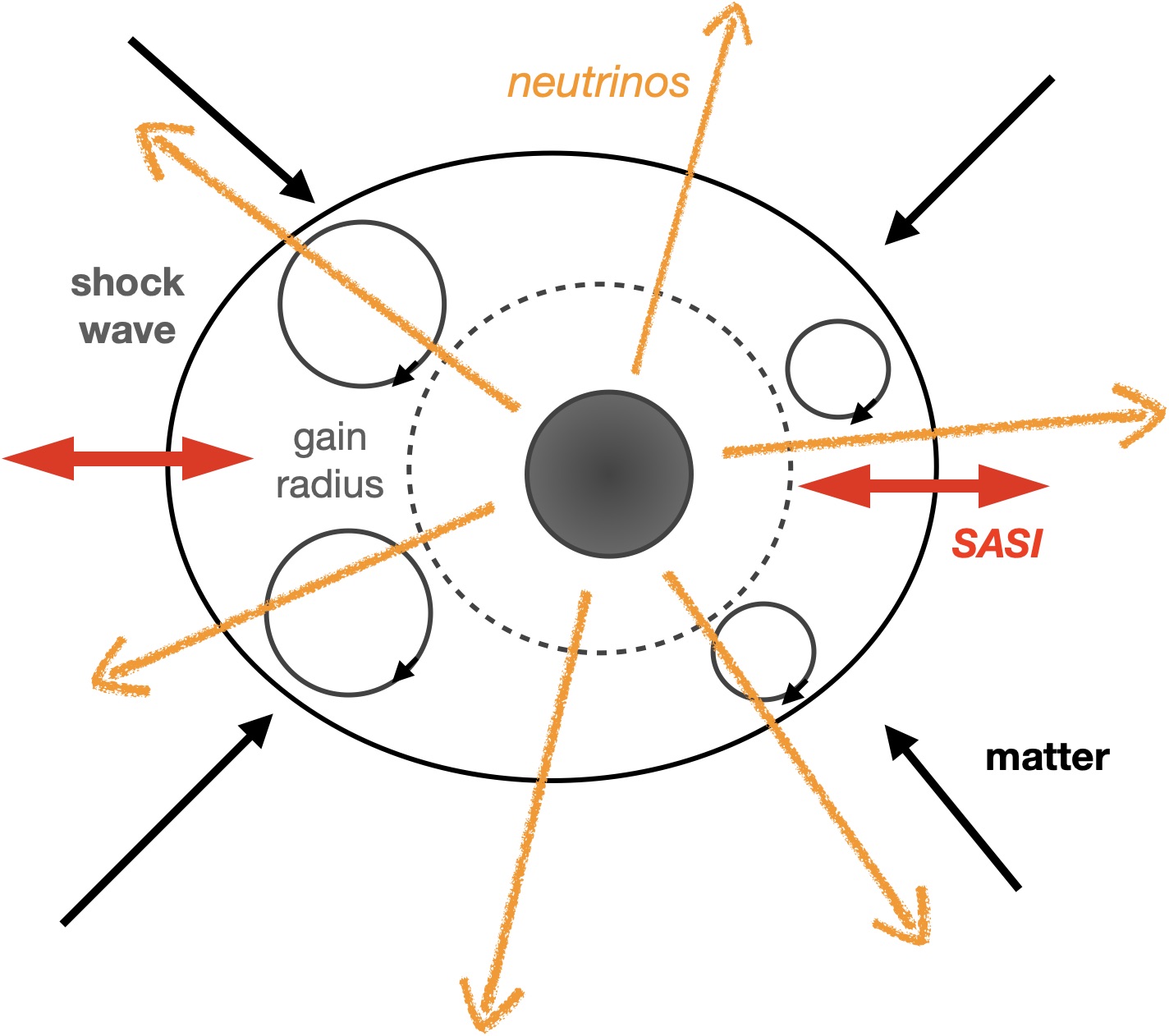}
\caption{\label{label}Schematic drawing of the inner regions of a core-collapse supernova. Outside the newly formed proto-neutron star that cools under neutrino emission, the gain region, surrounded by the region where neutrinos contribute to heating matter behind the shock.}
\end{minipage}\hspace{2pc}%
\begin{minipage}{14pc}
\includegraphics[width=12pc]{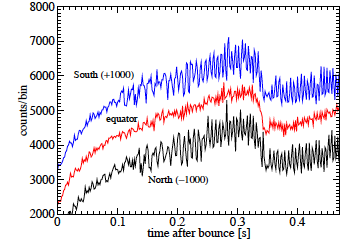}
\caption{\label{label}Time signals (plus background) in IceCube from a 25 M$_{\odot}$ supernova (10 kpc) for an observer located at the North, equatorial and South directions. The fast oscillations correspond to the Standing Accretion Shock Instability (SASI) \cite{Muller:2014rpb}.}
\end{minipage} 
\end{center}
\end{figure}
\end{center}

While most of the neutrinos are emitted during ten seconds, pre-supernova neutrinos might give advanced warning for a close supernova \cite{Kato:2020hlc}, 9 h before collapse with the current Super-Kamiokande+Gd phase for a star like Betelgeuse \cite{Super-Kamiokande:2022bwp}. Moreover late time neutrinos, emitted up to several tens of seconds, are not negligeable. An explosion at 10 kpc would produce 110 $\nu_e$ events in DUNE (up to 40 s), about 10 $\nu_x$ (and $\bar{\nu}_x$, $x = \mu, \tau$) in JUNO and 250 $\bar{\nu}_e$ in Super-Kamiokande \cite{Li:2020ujl}.

Neutrinos are also connected to the key open issue of where elements heavier than iron are synthetized (rapid neutron capture process or $r$-process). 
Supernovae and binary neutron star mergers are considered the main sites. While only the most energetic supernovae appear to provide suitable astrophysical conditions, a strong r-process can occur in the less frequent binary neutron star mergers. The unique event GW170817 gave indirect evidence for $r$-process elements in such sites \cite{LIGOScientific:2017vwq}.
The first gravitational waves from a binary neutron star merger were indeed detected in concomitance with a short gamma ray-burst and a kilonova. The comparison of the electromagnetic signal with models provided indirect evidence for the presence of actinides (rare elements plateau) and maybe lanthanides in the ejecta (see for example \cite{Cowperthwaite:2017dyu,Tanaka:2017qxj}). 

Neutrino flavor evolution influence $r$-process nucleosynthesis. This is mainly due to the induced spectral swappings which modify $\nu$ interaction rates  on neutrons and protons and therefore the electron fraction\footnote{Note that for $Y_e > 0.5$ there is no $r$-process.} $Y_e = (n_e - n_{\bar{e}})/(n_n + n_p)$ ($n_i$ $i=e, p, n$ are number densities) and the nucleosynthetic abundances in $r$-process networks. 
An example is given in Figure 4 where the impact of the matter-neutrino resonance is shown \cite{Malkus:2015mda}, a flavor mechanism due to a cancellation between neutrino interactions with matter and with $\nu$. It occurs in particular in binary neutron star mergers because the matter is neutron rich, giving an enhanced production of $\bar{\nu}_e$ over $\nu_e$.

How much flavor evolution does impact $r$-process nucleosynthesis is still an open question. The numerical challenge is to determine self-consistently the evolution of the matter composition and of neutrino flavor in detailed simulations of the astrophysical settings.

\section{Flavor evolution in core-collapse supernovae}
How neutrino change flavor in dense environments is, as for now, difficult to answer in a conclusive way. The difficulty is inherent to the high dimensionality of the problem\footnote{The problem to solve is 7-dimensional (it depends on time $t$, space $\vec{x}$ and momentum $\vec{p}$).} and the fact that we are facing a non-linear many-body problem due to the presence of 
sizable neutral current $\nu\nu$ interactions as pointed out by Pantaleone long ago \cite{Pantaleone:1992eq}. In most studies, one solves the Liouville-Von-Neumann equations of motion for single particle density matrices  ($\hbar=c=1$) \cite{Sigl:1993ctk,Volpe:2013uxl,Serreau:2014cfa}

\begin{equation}
i (\partial_t +  \mathbf{v} \cdot \nabla ) \varrho_\mathbf{p}  = [h_\mathbf{p}, \varrho_\mathbf{p} ]      ~~~~~~~~ h_\mathbf{p} = h_{\rm vac} + h_{\rm mat} + h_{\nu\nu,  \mathbf{p}} 
\end{equation}
where $h$ is the mean-field Hamiltonian. It includes the vacuum term $h_{vac} = \frac{\mathsf{M}^2}{2E_\nu}$ that depends on the mixings and mass-squared differences ($E_\nu = \vert \mathbf{p} \vert$ and $\mathbf{v} = \mathbf{p}/E_{\nu }$ are the neutrino energy and velocity respectively),  $h_{\rm mat} = \sqrt{2} G_F n_e$ depends on the neutrino-matter potential ($G_F$ is the Fermi coupling constant), $h_{\nu\nu}$ is the $\nu\nu$ interaction term. The single particle density matrix is given by expectation values $\varrho_{ij} = \langle a^{\dagger}_j a_i \rangle$ with $i,j = 1, 2, ..N$ the flavor (or mass) indices (N is the number of neutrino families), where $a, a^{\dagger}$ are the creation and annihilation operators that satisfy the canonical anticommutation rules. A similar equation holds for antineutrinos with $\bar{\varrho } _{ij} = \langle b^{\dagger}_i b_j \rangle$  ($b, b^{\dagger}$ being the antiparticles creation and annihilation operators).
\begin{figure}
\begin{center}
\includegraphics[scale=0.7]{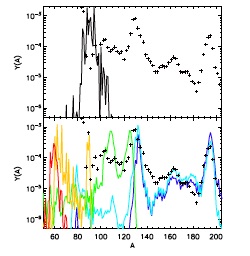}
\end{center}
\caption{\label{label}  Abundances as a function of the elements mass number A, in comparison with scaled solar mass residuals (black pluses). Upper figure: results with no flavor modification. Lower figure: results with the matter-neutrino resonance (see text) and different $\nu_{\mu}, \nu_{\tau}$ contributions, i.e. 0$\%$ (dark blue), 5 $\%$ (light blue), 10 $\%$ (green), 15 $\%$ (yellow) and 65 $\%$ (red) \cite{Malkus:2015mda}.}
\end{figure}

Several studies have investigated the validity of the mean-field approximation, including corrections at the mean-field level such as pairing correlations \cite{Volpe:2013uxl} and spin \cite{Vlasenko:2013fja} or helicity \cite{Serreau:2014cfa,Chatelain:2016xva} coherence. The use of algebraic methods and the Bethe {\it ansatz} have opened the possibility for an exact solution of the many-body problem of neutrino propagation in dense environments (without collisions) \cite{Pehlivan:2011hp}.
Contributions from collisions in schematic models with reduced dimensionality start being available \cite{Capozzi:2018clo,Richers:2019grc,Martin:2021xyl}. Note that, the first calculations of the Boltzmann equation, in the cosmological context, with the full collision term has been recently performed (including the mixings and the mean-field terms), contributing to the precise value of $N_{eff} = 3.0440$, for the effective number of degrees of freedom  \cite{Froustey:2020mcq}  (see \cite{Volpe:2015rla} for a review on the neutrino evolution equations). 

Concerning flavor mechanisms, for the well established Mikheev-Smirnov-Wolfenstein (MSW) effect \cite{Wolfenstein:1977ue,Mikheev:1986gs} only the evolution of neutrinos through the H-resonance is not fully determined because of the unknown sign of $\Delta m^2_{23}$ (for $\Delta m^2_{23} > 0$ the ordering is normal, inverted in the opposite case). Although supernova neutrinos could inform us about the neutrino mass ordering, as we will discuss, this will certainly be determined  in the coming future by experiments such as DUNE \cite{DUNE:2015lol}, JUNO or Hyper-K. The latter will measure the mass ordering  at about 3 $\sigma$ after 6 \cite{JUNO:2015zny} or 10 years \cite{Abe:2011ts} respectively. As for the shock waves and turbulence, their effects are understood in many respects (see \cite{Duan:2009cd} for a review). 

On the contrary, the impact of neutrino-neutrino interactions is still an open issue.  It is intensively studied since the work of Carlson et al \cite{Duan:2006an} that proposed the {\it bulb} model
and identified collective large scale modes, at about $\mathcal{O}(10^2$-$10^3)$ km from the neutrinosphere, nowadays called {\it slow}. {\it Fast} modes, uncovered by Sawyer \cite{Sawyer:2005jk} contrast with slow modes since they have short scales (meters or less) and occur behind the shock, very close to the neutrinosphere. Identified in detailed three-dimensional supernova simulations (see for example \cite{Abbar:2019zoq}) their influence on the supernova dynamics is not clear yet (for a review on fast modes see \cite{Tamborra:2020cul} and  \cite{Duan:2010bg,Mirizzi:2015eza} for slow modes).

\begin{figure}
\begin{center}
\includegraphics[scale=0.3]{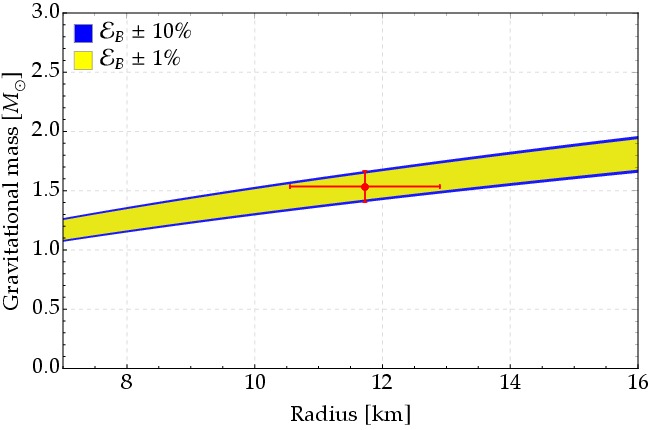}
\end{center}
\caption{\label{label} Reconstruction of the mass and radius of a newly formed neutron star during a supernova explosion at 10 kpc in Super-Kamiokande (yellow band) and Hyper-Kamiokande (blue band). The error is dominated by the uncertainty in the neutron star equation of state (EOS) (Lattimer and Prakash relation was used). The red cross corresponds to a star like SN1987A (see \cite{GalloRosso:2017hbp}).}
\end{figure}

\section{Future supernova neutrino observations}
Among the properties constrained by SN1987A events is the neutrino speed. The optical brightening followed neutrino emission by a few hours giving 
$\vert c - c_{\nu} \vert /c < 2 \times 10^{-9}$ \cite{Longo:1987ub}. The events also gave interesting limits on the neutrino magnetic moment\footnote{Neutrinos acquire a neutrino magnetic moment from effective one-photon coupling and quantum loops that give the standard model tiny value of $\mu_{\nu} = 3.2 \times 10^{-19} (m_{\nu}/{\rm eV})   \mu_B $ (see \cite{Giunti:2014ixa} for a review on neutrino electromagnetic properties).}, i.e. $\mu_{\nu} < 1.5 - 5 \times 10^{-11} \mu_B $ \cite{Lattimer:1988mf}.  A limit of  $\mu_{\nu} \sim 10^{-12} \mu_B $ could be obtained with a spherical gaseous TPC and a very intense radioactive source of 200 MCurie \cite{Giomataris:2003bp}. A network of such detectors (with 100 eV energy threshold) would be a dedicated long term supernova neutrino observatory \cite{Vergados:2016ueq}.

Operating since 2005, the Supernova Early Warning System is a network of detectors, based on different technologies (Cherenkov, scintillator, argon) which will observe the lucky event of the next (extra)galactic supernova. From several hundreds up to $10^6$ events will be detected (supernova at nominal distance of 10 kpc). Neutrino flavor and time
signal will be measured through inverse-beta decay, neutral current scattering on electrons and protons as well as neutrino interactions on nuclear targets, such as argon, carbon or oxygen which have a specific sensitivity to $\nu_e$ (see for example \cite{SNEWS:2022flw}). The upgraded SNEWS 2.0 will have the capability to point to the supernova via triangulation through its neutrinos which has tight timing requirements \cite{SNEWS:2020tbu}. 
\begin{center}
\begin{figure}[h]
\begin{center}
\begin{minipage}{14pc}
\includegraphics[width=14pc]{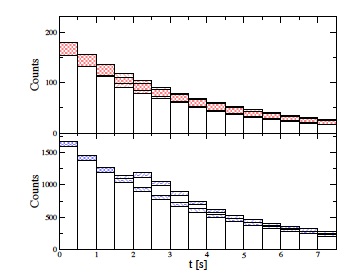}
\caption{\label{SK} Positron time signal in Super-Kamiokande from a supernova at 10 kpc. Energy bins (0.5 s) correspond to 10-19 MeV (upper) and above 20 MeV (lower figure). Results for an exponential cooling are also given to show the deviation due to the passage of the shock wave in the MSW region \cite{Gava:2009pj}.}
\end{minipage}\hspace{2pc}
\begin{minipage}{14pc}
\includegraphics[width=14pc]{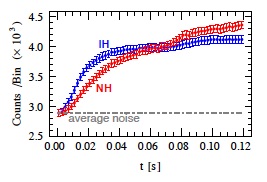}
\caption{\label{MH} Binned time signal (0.2 ms) with Poisson error estimate for the signal plus background noise in IceCube (the average noise is shown as a dot-dashed curve) \cite{Serpico:2011ir}.} 
\end{minipage} 
\end{center}
\end{figure}
\end{center}

Measurements of $\nu$-nucleus interaction cross sections are finally planned at SNS by the COHERENT Collaboration which performed the first detection of neutrino-nucleus coherent scattering, searched for many decades \cite{COHERENT:2017ipa}, providing among others new constraints on non-standard $\nu$-matter interactions.  
In particular, measurements are ongoing/planned on $^{208}$Pb,  $^{40}$Ar (CC and NC), $^{127}$I and $^{56}$Fe $^{16}$O $^{76}$Ge \cite{Kate,Barbeau:2021exu}. These measurements are important since they will provide a more precise knowledge of the nuclear spin and isospin response to neutrinos. They might also provide us with information on the (possible) quenching of the axial vector coupling constant $g_A$ for forbidden Gamow-Teller (type) transitions in atomic nuclei \cite{Volpe:2005iy}.

Clearly, the detection of supernova neutrinos from the next (extra)galactic supernova is crucial both for astrophysics and for particle physics. 
Besides the previously discussed explosion mechanism, one will determine the total gravitational binding energy with a precision of about 11 $\%$ (3 $\%$) percent\footnote{These results comes from a 9+1 degrees of freedom likelihood analysis where the parameters characterizing the neutrino fluxes are left free to vary (within the priors).} with Super-Kamiokande (Hyper-Kamkiokande \cite{GalloRosso:2017mdz}). This would give the mass-radius relation of the newly formed neutron star \cite{GalloRosso:2017hbp} for which the neutron star EOS would represent the main uncertainty, as seen from Figure 5. 

\begin{center}
\begin{figure}[h]
\begin{center}
\begin{minipage}{14pc}
\includegraphics[width=14pc]{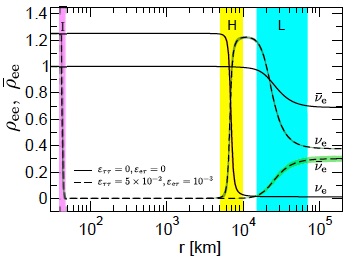}
\caption{\label{NSI-I}  NSI effects in a core-collapse supernova. Close to the neutrinosphere is the I-resonance, due to standard and non-standard neutrino matter interactions. The MSW H- and L-resonance are also shown at further distances \cite{Esteban-Pretel:2007zkv}.}
\end{minipage}\hspace{2pc}
\begin{minipage}{14pc}
\includegraphics[width=14pc]{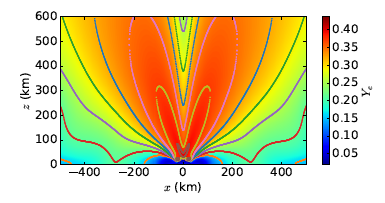}
\caption{\label{NSI-BNS} NSI effects in a binary neutron star merger remnant (in the center). The full lines show the locations of the I-resonance in the funnel and the polar regions. The colors correspond to the electron fraction $Y_e$  \cite{Chatelain:2017yxx}.} 
\end{minipage} 
\end{center}
\end{figure}
\end{center}
Obviously, the supernova signal is intertwined with unknown neutrino properties. There are currently hints for a non-zero Dirac CP violating phase and for normal mass ordering.
Balantekin, Gava and Volpe showed that, in supernovae, if $\delta \neq 0$ there can be CP violating effects on the electron flavored fluxes as well  \cite{Balantekin:2007es}, contrarily to what was previously believed. The impact found on the fluxes was small, unless new physics introduces differences between the $\nu_{\mu}, \nu_{\tau}$ fluxes. The combined effect of a non-zero Majorana phases in presence of strong magnetic fields can introduce supplementary resonances \cite{Popov:2021icg}. 

The identification of the neutrino mass ordering using the next supernova was extensively discussed (see for example Figures 6 and 7) . One possibility is offered by the passage of the shock wave in the MSW region during the explosion. This produces multiple MSW resonances and dips and bumps in the time signal, depending on neutrino energy, compared to an exponential cooling. These features are expected in the $\nu_e$ detection channel (normal mass ordering) in a detector like DUNE, or in the $\bar{\nu}_e$ detection channel (inverted) in Cherenkov or scintillator detectors such as Super-Kamiokande (figure 6), Hyper-Kamiokande or JUNO.

Non-standard neutrino-matter interactions (NSI) influence the neutrino flavor content in core-collapse supernovae \cite{Stapleford:2016jgz} and in remnants of binary neutron star mergers \cite{Chatelain:2017yxx}. Upper limits on non-standard interactions are obtained from solar, oscillation experiments \cite{Farzan:2017xzy} and coherent neutrino-nucleus scattering \cite{COHERENT:2017ipa}.  NSI can introduce a significant modification of the neutrino flavor content, due e.g. to an MSW-like phenomenon called the I-resonance which is due to a cancellation between the standard and non-standard matter terms \cite{Esteban-Pretel:2007zkv}, but can  behave as a synchronized MSW in presence of sizeable $\nu\nu$ interactions \cite{Chatelain:2017yxx}  (Figures 8 and 9). Interestingly, numerical simulations show that even small NSI couplings, far from the upper bounds, induce an interplay of flavor phenomena \cite{Stapleford:2016jgz,Chatelain:2017yxx}. 

\begin{figure}
\begin{center}
\includegraphics[scale=0.35]{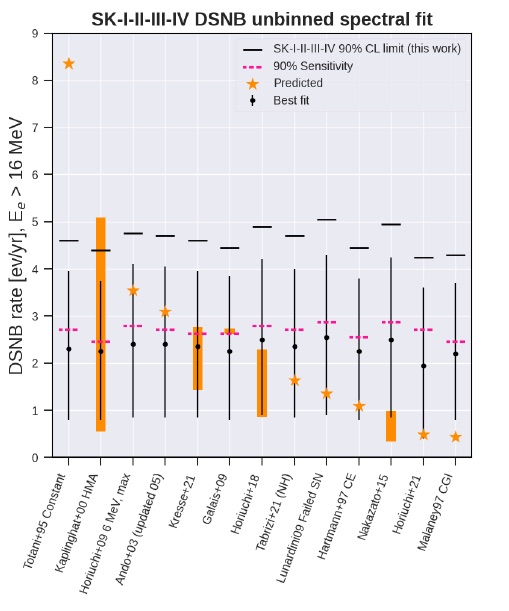}
\end{center}
\caption{\label{label} Expected DSNB rates (in orange, as a range or one value) according to different models and 90 $\%$ C.L. upper limits, best-fit values (1$\sigma$) and expected sensitivities from SK I to SK IV data \cite{Super-Kamiokande:2021jaq}.}
\end{figure}

Still unobserved is the diffuse supernova neutrino background (DSNB) made of the neutrinos produced by past supernovae. At present the analysis of Super-Kamiokande data gives a positive indication - a statistical fluctuation over background at 1.5$\sigma$  \cite{Super-Kamiokande:2021jaq}. The redshifted energies of the DSNB, covering an energy range similar to the core-collapse supernova one\footnote{Note that mostly redshifts $z = 0, 1, 2$ contribute.}, offer  a detection window, typically between 10 MeV and 30 MeV.
At low energies solar $\nu_e$ or reactor $\bar{\nu}_e$ overwhelm the DSNB, at higher energies atmospheric backgrounds\footnote{Note that neutral current atmospheric events require careful attention even in the DSNB detection window \cite{Priya:2017bmm,Super-Kamiokande:2021jaq}.}. Predictions for the DSNB rates and expected sensitivities from SK I-IV data are shown in Figure 10,
showing that results from 4 models are on par with SK data. 

Physics-wise the DSNB observation has complementary features to the one of the next supernova. 
In fact, the DSNB fluxes depend on the core-collapse supernova rate, on the fraction of failed supernovae (see \cite{Lunardini:2009ya} for a review) and from binary interactions \cite{Kresse:2020nto,Horiuchi:2020jnc} which are still uncertain. The relic fluxes are also sensitive to fundamental unknown properties such as neutrino decay (see for example \cite{DeGouvea:2020ang}) and to flavor evolution e.g. shock wave effects that can influence the rates as much as the MSW effect, as shown in \cite{Galais:2009wi}. 
We are entering an exciting phase since the discovery of the DSNB is expected from the running Super-Kamiokande+Gadolinium  experiment \cite{Super-Kamiokande:2021jaq} and the upcoming JUNO \cite{JUNO:2022lpc} and Hyper-Kamiokande \cite{Abe:2011ts}.

Definitely, neutrino astrophysics keeps bringing discoveries and surprises. Theoretically, our understanding of neutrino flavor evolution in dense environments has made great progress in the last fifteen years, but many challenges are still ahead. The future observations of a supernova and of the diffuse supernova neutrino background will allow major steps forward in this domain.

\end{document}